\documentclass[12pt]{article}

\renewcommand {\theequation}{\arabic{section}.\arabic{equation}}

\newcommand{\Tr}{\mathop{\rm Tr}\nolimits}

\title{CALCULATION OF THE SCATTERING AMPLITUDES}

\author {A.L.~Bondarev%
\thanks{\rm e-mail: bondarev@hep.by} \\ {\sl NC PHEP BSU,
Minsk, Belarus}}

\date{}

\begin{document}
\maketitle


\section {General equation for the calculation of amplitudes}
\setcounter{equation}{0}

It is well known that the high order calculations of the
observables within  the perturbation theory has the serious
difficulties (especially, if we take into account the polarization
effects). Method of the direct calculation of amplitudes
simplifies these calculations essentially.

There is an even number $(2N)$ of fermions in initial and final
state for any reaction with Dirac particles. Therefore every
diagram contains $N$ non-closed fermion lines. The expression
\begin{equation}
\displaystyle
 M_{if} = \bar{u}_f Q u_i = \Tr ( Q u_i \bar{u}_f )
\label{e1.1}
\end{equation}
corresponds to every line in the amplitude of the process, $\,
u_i$, $u_f$ are the Dirac bispinors for free particles and $Q$ is
a  matrix operator characterizing the interaction. The operator
$Q$ is expressed as a linear combination of the products of the
Dirac $\gamma$-matrices (or their contractions with four-vectors)
and can have any number of free Lorentz indexes.

We must obtain the expression for the operator
\begin{equation}
\displaystyle
u_i \bar{u}_f
\label{e1.2}
\end{equation}
in (\ref{e1.1}).

Note that Dirac equation
\begin{equation}
\displaystyle
{\hat p} u(p,n) = m u(p,n) \; ,
\label{e1.3}
\end{equation}
and equation for the axis of the spin projections
\begin{equation}
\displaystyle
\gamma^5 {\hat n} u(p,n) =  \pm u(p,n)
\label{e1.4}
\end{equation}
as well as normalization condition for bispinor define $u(p,n)$ up
to the phase factor ($p$ is four-momentum, $n$ is  the four-vector
specifying the axis of the spin projections and
$\displaystyle \hat{a} = \gamma_{\mu} a^{\mu} \; $
for any four-vector $a$)
\footnote{ We use the same metric as in the book \cite{r1.1}:
$$ \displaystyle a^{\mu} = ( a_0 , \vec{a} ) , \;\;\; a_{\mu} = (
a_0, -\vec{a} ) , \;\;\; ab = a_{\mu} b^{\mu} = a_0 b_0 - \vec{a}
\vec{b} , \;\;\; \gamma^5 = {\it i} \gamma^0 \gamma^1 \gamma^2
\gamma^3 \; . $$
}.

Let use this fact to construct the operator (\ref{e1.2}):
\begin{equation}
\displaystyle
u_i \simeq u_i { \bar{u}_i u \over | \bar{u}_i u | } = { {\cal
P}_i  \over \sqrt { \Tr ( {\cal P} {\cal P}_i ) } } u  \; ,
\label{e1.5}
\end{equation}
\begin{equation}
\displaystyle
\bar{u}_f \simeq { \bar{u} u_f \over | \bar{u} u_f | } \bar{u}_f =
\bar{u} { {\cal P}_f  \over \sqrt { \Tr ( {\cal P} {\cal P}_f ) }
}  \; ,
\label{e1.6}
\end{equation}
where $u$ is an arbitrary bispinor and ${\cal P}$ are projection
operators for Dirac particles.
\begin{equation}
\displaystyle
{\cal P}(p,n) = u(p,n) \bar{u}(p,n) = { 1 \over 2 } ( \hat{p} + m
)                      ( 1 + \gamma^5 \hat{n} )
\label{e1.7}
\end{equation}
is the projection operator for a particle with mass $m$,
$$ \displaystyle p^2 = m^2 \; , \;\; n^2 = -1 \; , \;\; pn = 0 \;
, \;\; \bar{u} u = 2 m \; .$$
For massless particle the projection operator has form:
\begin{equation}
\displaystyle
u_{\pm}(p) \bar{u}_{\pm}(p) = { 1 \over 2 }( 1 \pm \gamma^5 )
\hat{p} = {\cal P}_{\pm}(p)
\label{e1.8}
\end{equation}
where:
$$ \displaystyle p^2 = 0 \; , \;\; \bar{u}_{\pm} \gamma_{\mu}
u_{\pm} =  2 p_{\mu} $$
(signs $\pm $ correspond to the helicity of particle) .

As a result we have
\begin{equation}
\displaystyle
u_i \bar{u}_f \simeq  { {\cal P}_i {\cal P} {\cal P}_f \over \sqrt
{ \Tr ( {\cal P} {\cal P}_i ) \Tr ( {\cal P} {\cal P}_f ) } } \; .
\label{e1.9}
\end{equation}
It follows from (\ref{e1.5}), (\ref{e1.6}) that both $u_i$ and
$\bar{u}_f$ obtain individual phase factors, i.e. direct as well
as exchange diagrams multiplied by the same phase factor, and
consequently one can ignore it. Therefore we will use the equality
sign instead  of the symbol $ \simeq $ in formulae for amplitudes.

Finally, we have the general formula for the calculation of
amplitudes
\begin{equation}
\displaystyle
\bar{u}_f Q u_i = { \Tr ( Q {\cal P}_i {\cal P} {\cal P}_f ) \over
\sqrt { \Tr ( {\cal P} {\cal P}_i ) \Tr ( {\cal P} {\cal P}_f ) }
}  \; .
\label{e1.10}
\end{equation}
This expression enables to calculate the amplitude numerically.
The complex values obtained in this way may be used to evaluate
the squared amplitude.

Note that this method can be generalized easily to the case of the
reaction with antiparticles. To do this it is sufficient in
(\ref{e1.10}) to replace the projection operators of particles by
the operators of antiparticles.

More details may be found in \cite{r1.2}.


\section {Example: the calculation of electron-electron scattering amplitudes}
\setcounter{equation}{0}

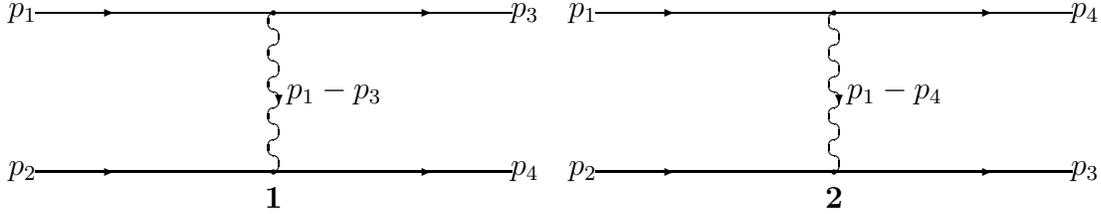
\begin{figure}[ht]
\begin{tabular}{cc}
\begin{picture}(200,80)
\put(05,10){\makebox(0,0){$ p_2 $}}
\put(05,70){\makebox(0,0){$ p_1 $}}
\put(195,10){\makebox(0,0){$ p_4 $}}
\put(195,70){\makebox(0,0){$ p_3 $}}
\put(123,40){\makebox(0,0){$ p_1 - p_3 $}}
\put(100,0){\makebox(0,0){$ {\bf 1} $}}
%
\put(10,10){\line(1,0){180.}}
\put(10,70){\line(1,0){180.}}
\put(100,10){\circle*{2}}
\put(100,70){\circle*{2}}
\put(40,10){\vector(1,0){0.}}
\put(40,70){\vector(1,0){0.}}
\put(160,10){\vector(1,0){0.}}
\put(160,70){\vector(1,0){0.}}
\put(102,35){\vector(0,-1){0.}}
\multiput(100,13)(0,12){5}{\oval(4.0,6.0)[r]}
\multiput(100,19)(0,12){5}{\oval(4.0,6.0)[l]}
\end{picture}
&
\begin{picture}(200,80)
\put(05,10){\makebox(0,0){$ p_2 $}}
\put(05,70){\makebox(0,0){$ p_1 $}}
\put(195,10){\makebox(0,0){$ p_3 $}}
\put(195,70){\makebox(0,0){$ p_4 $}}
\put(123,40){\makebox(0,0){$ p_1 - p_4 $}}
\put(100,0){\makebox(0,0){$ {\bf 2} $}}
%
\put(10,10){\line(1,0){180.}}
\put(10,70){\line(1,0){180.}}
\put(100,10){\circle*{2}}
\put(100,70){\circle*{2}}
\put(40,10){\vector(1,0){0.}}
\put(40,70){\vector(1,0){0.}}
\put(160,10){\vector(1,0){0.}}
\put(160,70){\vector(1,0){0.}}
\put(102,35){\vector(0,-1){0.}}
\multiput(100,13)(0,12){5}{\oval(4.0,6.0)[r]}
\multiput(100,19)(0,12){5}{\oval(4.0,6.0)[l]}
\end{picture}
\end{tabular}
\caption{Diagrams for electron-electron scattering} \label{Fg1}
\end{figure}

As an illustration of application of the method we will consider
the lowest order amplitudes of electron-electron scattering. Two
Feynman diagrams presented in Fig.\ref{Fg1} correspond to this
process.
%
%
\begin{equation}
\displaystyle
M_1 = {1 \over (p_1-p_3)^2} \, \bar{u}_3 \gamma_{\rho} u_1 \cdot
\bar{u}_4 \gamma^{\rho} u_2 \, ,
\label{e2.1}
\end{equation}
\begin{equation}
\displaystyle
M_2 = - {1 \over (p_1-p_4)^2} \, \bar{u}_4 \gamma_{\rho} u_1 \cdot
\bar{u}_3 \gamma^{\rho} u_2 \, .
\label{e2.2}
\end{equation}

There are two variants of formula (\ref{e1.10}) for massive Dirac
particles:
\begin{equation}
\displaystyle
\bar{u}_f Q u_i = { \Tr \left[ ( 1 - {\gamma}^5 ) \hat{q}  (
\hat{p}_f + m \hat{n}_f + m - \hat{p}_f \hat{n}_f ) Q (\hat{p}_i +
m \hat{n}_i + m + \hat{p}_i \hat{n}_i ) \right] \over 8 \sqrt{  (
q p_i ) + m ( q n_i ) }  \sqrt{ ( q p_f ) + m ( q n_f ) } } \;\; ,
\label{e2.3}
\end{equation}
if
$ \displaystyle
{\cal P} = {1 \over 2} ( 1 - {\gamma}^5 ) \hat{q}  \;\; ,
$
(the massless four-vector $q$ can be chosen arbitrary, but it has
to be the same for all non-closed fermion lines of the considered
diagrams), and
\begin{equation}
\displaystyle
\left( \bar{u}_f Q u_i \right)' = { \Tr \left[ ( 1 + {\gamma}^5 )
\hat{q} ( \hat{p}_f - m \hat{n}_f + m + \hat{p}_f \hat{n}_f ) Q
(\hat{p}_i - m \hat{n}_i + m - \hat{p}_i \hat{n}_i ) \right] \over
8 \sqrt{ ( q p_i ) - m ( q n_i ) }  \sqrt{ ( q p_f ) - m ( q n_f )
} } \;\; ,
\label{e2.4}
\end{equation}
if
$ \displaystyle {\cal P} = {1 \over 2} ( 1 + {\gamma}^5 ) \hat{q}
\;\; .$
Relation between (\ref{e2.3}) and (\ref{e2.4}) gives by
\begin{equation}
\displaystyle
\bar{u}_f Q u_i  = \left( \bar{u}_f Q u_i \right)' \cdot {- \Tr
\left[ ( 1 + {\gamma}^5 ) \hat{q} \hat{p}_i \hat{n}_i \hat{q}
\hat{p}_f \hat{n}_f \right] \over 8 \sqrt{ ( q p_i )^2 - m^2 ( q
n_i )^2 }  \sqrt{ ( q p_f )^2 - m^2 ( q n_f )^2  } } \;\; .
\label{e2.5}
\end{equation}
For definiteness we will use (\ref{e2.3}).

Note, to check formula (\ref{e2.5}) it necessary to use identity:
\begin{equation}
\displaystyle
( 1 \pm {\gamma}^5 ) \hat{q} Q ( 1 \pm {\gamma}^5 ) \hat{q} = \Tr
\left[ ( 1 \pm {\gamma}^5 ) \hat{q} Q \right] ( 1 \pm {\gamma}^5 )
\hat{q} \;\; ,
\label{e2.6}
\end{equation}
where an operator $Q$ is any matrix. (The proof of (\ref{e2.6}) is
given in \cite{r1.2}.) Besides, we will use well-known formulae of
$\gamma$-matrix algebra:
\begin{equation}
\displaystyle
\gamma_{\rho} Q^{2n+1} \gamma^{\rho} = -2Q^{2n+1}_R
\label{e2.7}
\end{equation}
and
\begin{equation}
\displaystyle
(1 \pm {\gamma}^5) \gamma_{\rho} Q^{2n} \gamma^{\rho} = (1 \pm
{\gamma}^5) \Tr \left[(1 \mp {\gamma}^5) Q^{2n}  \right] \; ,
\label{e2.8}
\end{equation}
where $\displaystyle Q^{2n+1}$ is a product of odd number of the
$\gamma$-matrices; $\displaystyle Q^{2n+1}_R$ is a product of the
same $\gamma$-matrices, rewritten in reversal order and
$\displaystyle Q^{2n}$ is a product of even number of the
$\gamma$-matrices.

Thus, we obtain from (\ref{e2.3}):
\begin{equation}
\displaystyle
\bar{u}_3 \gamma_{\rho} u_1 = { \Tr \left[ ( 1 - {\gamma}^5 )
\hat{q} ( \hat{p}_3 + m \hat{n}_3 + m - \hat{p}_3 \hat{n}_3 )
\gamma_{\rho} (\hat{p}_1 + m \hat{n}_1 + m + \hat{p}_1 \hat{n}_1 )
\right] \over 8 \sqrt{ ( q p_1 ) + m ( q n_1 ) }  \sqrt{ ( q p_3 )
+ m ( q n_3 ) } } \; ,
\label{e2.9}
\end{equation}
\begin{equation}
\displaystyle
\bar{u}_4 \gamma^{\rho} u_2 = { \Tr \left[ ( 1 - {\gamma}^5 )
\hat{q} ( \hat{p}_4 + m \hat{n}_4 + m - \hat{p}_4 \hat{n}_4 )
\gamma^{\rho} (\hat{p}_2 + m \hat{n}_2 + m + \hat{p}_2 \hat{n}_2 )
\right] \over 8 \sqrt{ ( q p_2 ) + m ( q n_2 ) }  \sqrt{ ( q p_4 )
+ m ( q n_4 ) } } \; ,
\label{e2.10}
\end{equation}
\begin{equation}
\begin{array}{r} \displaystyle
\bar{u}_3 \gamma_{\rho} u_1 \cdot \bar{u}_4 \gamma^{\rho} u_2 = {1
\over  8^2 \sqrt{ (q p_1) + m (q n_1) }  \sqrt{ (q p_2) + m (q
n_2) } \sqrt{ (q p_3) + m (q n_3) } \sqrt{ (q p_4) + m (q n_4) } }
\cdot
                \\[0.8cm]    \displaystyle
\Big\{ \Tr \left[ (1 - {\gamma}^5) \hat{q} (\hat{p}_3 + m
\hat{n}_3) \gamma_{\rho} (\hat{p}_1 + m \hat{n}_1) \right] \Tr
\left[ (1 - {\gamma}^5) \hat{q} (\hat{p}_4 + m \hat{n}_4)
\gamma^{\rho} (\hat{p}_2 + m \hat{n}_2) \right]
               \\[0.5cm]    \displaystyle
+ \Tr \left[ (1 - {\gamma}^5) \hat{q} (\hat{p}_3 + m \hat{n}_3)
\gamma_{\rho} (\hat{p}_1 + m \hat{n}_1) \right] \Tr \left[ (1 -
{\gamma}^5) \hat{q} (m - \hat{p}_4 \hat{n}_4) \gamma^{\rho} (m +
\hat{p}_2 \hat{n}_2) \right]
               \\[0.5cm]    \displaystyle
+ \Tr \left[ (1 - {\gamma}^5) \hat{q} (m - \hat{p}_3 \hat{n}_3)
\gamma_{\rho} (m + \hat{p}_1 \hat{n}_1) \right] \Tr \left[ (1 -
{\gamma}^5) \hat{q} (\hat{p}_4 + m \hat{n}_4) \gamma^{\rho}
(\hat{p}_2 + m \hat{n}_2) \right]
              \\[0.5cm]    \displaystyle
+ \Tr \left[ (1 - {\gamma}^5) \hat{q} (m - \hat{p}_3 \hat{n}_3)
\gamma_{\rho} (m + \hat{p}_1 \hat{n}_1) \right] \Tr \left[ (1 -
{\gamma}^5) \hat{q} (m - \hat{p}_4 \hat{n}_4) \gamma^{\rho} (m +
\hat{p}_2 \hat{n}_2) \right]
 \Big\} \; .
\end{array}
\label{e2.11}
\end{equation}
Using (\ref{e2.6}), (\ref{e2.7}), for first term in (\ref{e2.11})
we have:
\begin{equation}
\begin{array}{r} \displaystyle
\Tr \left[ (1 - {\gamma}^5) \hat{q} (\hat{p}_3 + m \hat{n}_3)
\gamma_{\rho} (\hat{p}_1 + m \hat{n}_1) \right] \Tr \left[ (1 -
{\gamma}^5) \hat{q} (\hat{p}_4 + m \hat{n}_4) \gamma^{\rho}
(\hat{p}_2 + m \hat{n}_2) \right]
               \\[0.5cm]    \displaystyle
= 2 \Tr \left[ (1 - {\gamma}^5) \hat{q} (\hat{p}_3 + m \hat{n}_3)
\gamma_{\rho} (\hat{p}_1 + m \hat{n}_1) \hat{q} (\hat{p}_4 + m
\hat{n}_4) \gamma^{\rho} (\hat{p}_2 + m \hat{n}_2) \right]
             \\[0.5cm]    \displaystyle
= -4 \Tr \left[ (1 - {\gamma}^5) \hat{q} (\hat{p}_3 + m \hat{n}_3)
(\hat{p}_4 + m \hat{n}_4) \hat{q} (\hat{p}_1 + m \hat{n}_1)
(\hat{p}_2 + m \hat{n}_2) \right]
 \; .
\end{array}
\label{e2.12}
\end{equation}
For second term in (\ref{e2.11}) we obtain through (\ref{e2.6}),
(\ref{e2.8})
\begin{equation}
\begin{array}{r} \displaystyle
\Tr \left[ (1 - {\gamma}^5) \hat{q} (\hat{p}_3 + m \hat{n}_3)
\gamma_{\rho} (\hat{p}_1 + m \hat{n}_1) \right] \Tr \left[ (1 -
{\gamma}^5) \hat{q} (m - \hat{p}_4 \hat{n}_4) \gamma^{\rho} (m +
\hat{p}_2 \hat{n}_2) \right]
               \\[0.5cm]    \displaystyle
= 2 \Tr \left[ (1 - {\gamma}^5) \hat{q} (\hat{p}_3 + m \hat{n}_3)
\gamma_{\rho} (\hat{p}_1 + m \hat{n}_1) \hat{q} (m - \hat{p}_4
\hat{n}_4) \gamma^{\rho} (m + \hat{p}_2 \hat{n}_2) \right]
             \\[0.5cm]    \displaystyle
= 2 \Tr \left[ (1 - {\gamma}^5) \hat{q} (\hat{p}_3 + m \hat{n}_3)
(m + \hat{p}_2 \hat{n}_2) \right] \Tr \left[ (1 - {\gamma}^5)
\hat{q} (m - \hat{p}_4 \hat{n}_4) (\hat{p}_1 + m \hat{n}_1)
\right] \; .
\end{array}
\label{e2.13}
\end{equation}
In the same way
\begin{equation}
\begin{array}{r} \displaystyle
\Tr \left[ (1 - {\gamma}^5) \hat{q} (m - \hat{p}_3 \hat{n}_3)
\gamma_{\rho} (m + \hat{p}_1 \hat{n}_1) \right] \Tr \left[ (1 -
{\gamma}^5) \hat{q} (\hat{p}_4 + m \hat{n}_4) \gamma^{\rho}
(\hat{p}_2 + m \hat{n}_2) \right]
             \\[0.5cm]    \displaystyle
= 2 \Tr \left[ (1 - {\gamma}^5) \hat{q} (\hat{p}_4 + m \hat{n}_4)
(m + \hat{p}_1 \hat{n}_1) \right] \Tr \left[ (1 - {\gamma}^5)
\hat{q} (m - \hat{p}_3 \hat{n}_3) (\hat{p}_2 + m \hat{n}_2)
\right] \; ,
\end{array}
\label{e2.14}
\end{equation}
and
\begin{equation}
\begin{array}{r} \displaystyle
\Tr \left[ (1 - {\gamma}^5) \hat{q} (m - \hat{p}_3 \hat{n}_3)
\gamma_{\rho} (m + \hat{p}_1 \hat{n}_1) \right] \Tr \left[ (1 -
{\gamma}^5) \hat{q} (m - \hat{p}_4 \hat{n}_4) \gamma^{\rho} (m +
\hat{p}_2 \hat{n}_2) \right]
             \\[0.5cm]    \displaystyle
= -4 \Tr \left[ (1 - {\gamma}^5) \hat{q} (m - \hat{p}_3 \hat{n}_3)
(m + \hat{p}_4 \hat{n}_4) \hat{q} (m - \hat{p}_1 \hat{n}_1) (m +
\hat{p}_2 \hat{n}_2) \right] \; .
\end{array}
\label{e2.15}
\end{equation}
As a result we have
\begin{equation}
\begin{array}{r} \displaystyle
\bar{u}_3 \gamma_{\rho} u_1 \cdot \bar{u}_4 \gamma^{\rho} u_2 = {1
\over  32 \sqrt{ (q p_1) + m (q n_1) }  \sqrt{ (q p_2) + m (q n_2)
} \sqrt{ (q p_3) + m (q n_3) } \sqrt{ (q p_4) + m (q n_4) } }
\cdot
                \\[0.8cm]    \displaystyle
\Big\{ -2 \Tr \left[ (1 - {\gamma}^5) \hat{q} (\hat{p}_3 + m
\hat{n}_3) (\hat{p}_4 + m \hat{n}_4) \hat{q} (\hat{p}_1 + m
\hat{n}_1) (\hat{p}_2 + m \hat{n}_2) \right]
               \\[0.5cm]    \displaystyle
+ \Tr \left[ (1 - {\gamma}^5) \hat{q} (\hat{p}_3 + m \hat{n}_3) (m
+ \hat{p}_2 \hat{n}_2) \right] \Tr \left[ (1 - {\gamma}^5) \hat{q}
(m - \hat{p}_4 \hat{n}_4) (\hat{p}_1 + m \hat{n}_1) \right]
               \\[0.5cm]    \displaystyle
+ \Tr \left[ (1 - {\gamma}^5) \hat{q} (\hat{p}_4 + m \hat{n}_4) (m
+ \hat{p}_1 \hat{n}_1) \right] \Tr \left[ (1 - {\gamma}^5) \hat{q}
(m - \hat{p}_3 \hat{n}_3) (\hat{p}_2 + m \hat{n}_2) \right]
              \\[0.5cm]    \displaystyle
-2 \Tr \left[ (1 - {\gamma}^5) \hat{q} (m - \hat{p}_3 \hat{n}_3)
(m + \hat{p}_4 \hat{n}_4) \hat{q} (m - \hat{p}_1 \hat{n}_1) (m +
\hat{p}_2 \hat{n}_2) \right]
 \Big\} \; .
\end{array}
\label{e2.16}
\end{equation}

In principle we may use (\ref{e2.16}) for calculation of amplitude
but it possible to simplify essentially the last term in this
expression. Really, using consequence of formula (\ref{e2.5})
\begin{equation}
\begin{array}{c} \displaystyle
\Tr \left[ (1 - {\gamma}^5) \hat{q} (m - \hat{p}_4 \hat{n}_4)
\gamma^{\rho} (m + \hat{p}_2 \hat{n}_2) \right]
              \\[0.5cm]    \displaystyle
= - \Tr \left[ (1 - {\gamma}^5) \hat{q} (\hat{p}_2 - m \hat{n}_2)
\gamma^{\rho} (\hat{p}_4 - m \hat{n}_4) \right] \cdot { \Tr \left[
( 1 - {\gamma}^5 ) \hat{q} \hat{p}_4 \hat{n}_4 \hat{q} \hat{p}_2
\hat{n}_2 \right] \over 8 \left[ (q p_2) - m (q n_2) \right] \cdot
\left[ (q p_4) - m (q n_4) \right] } \;\; ,
\end{array}
\label{e2.17}
\end{equation}
we have instead of (\ref{e2.15})
\begin{equation}
\begin{array}{r} \displaystyle
\Tr \left[ (1 - {\gamma}^5) \hat{q} (m - \hat{p}_3 \hat{n}_3)
\gamma_{\rho} (m + \hat{p}_1 \hat{n}_1) \right] \Tr \left[ (1 -
{\gamma}^5) \hat{q} (m - \hat{p}_4 \hat{n}_4) \gamma^{\rho} (m +
\hat{p}_2 \hat{n}_2) \right]
             \\[0.5cm]    \displaystyle
= -2 \Tr \left[ (1 - {\gamma}^5) \hat{q} (\hat{p}_2 - m \hat{n}_2)
(m + \hat{p}_1 \hat{n}_1) \right] \Tr \left[ (1 - {\gamma}^5)
\hat{q} (m - \hat{p}_3 \hat{n}_3) (\hat{p}_4 - m \hat{n}_4)
\right]
             \\[0.5cm]    \displaystyle
\cdot { \Tr \left[ ( 1 - {\gamma}^5 ) \hat{q} \hat{p}_4 \hat{n}_4
\hat{q} \hat{p}_2 \hat{n}_2 \right] \over 8 \left[ (q p_2) - m (q
n_2) \right] \cdot \left[ (q p_4) - m (q n_4)  \right] } \; .
\end{array}
\label{e2.18}
\end{equation}
Similarly, for first term of (\ref{e2.16}), using the other
consequence of formula (\ref{e2.5})
\begin{equation}
\begin{array}{c} \displaystyle
\Tr \left[ (1 - {\gamma}^5) \hat{q} (\hat{p}_4 + m \hat{n}_4)
\gamma^{\rho} (\hat{p}_2 + m \hat{n}_2) \right]
              \\[0.5cm]    \displaystyle
= - \Tr \left[ (1 - {\gamma}^5) \hat{q} (m + \hat{p}_2 \hat{n}_2)
\gamma^{\rho} (m - \hat{p}_4 \hat{n}_4) \right] \cdot { \Tr \left[
( 1 - {\gamma}^5 ) \hat{q} \hat{p}_4 \hat{n}_4 \hat{q} \hat{p}_2
\hat{n}_2 \right] \over 8 \left[ (q p_2) - m (q n_2) \right] \cdot
\left[ (q p_4) - m (q n_4) \right] } \;\; ,
\end{array}
\label{e2.19}
\end{equation}
we obtain instead of (\ref{e2.12})
\begin{equation}
\begin{array}{r} \displaystyle
\Tr \left[ (1 - {\gamma}^5) \hat{q} (\hat{p}_3 + m \hat{n}_3)
\gamma_{\rho} (\hat{p}_1 + m \hat{n}_1) \right] \Tr \left[ (1 -
{\gamma}^5) \hat{q} (\hat{p}_4 + m \hat{n}_4) \gamma^{\rho}
(\hat{p}_2 + m \hat{n}_2) \right]
               \\[0.5cm]    \displaystyle
= - 2 \Tr \left[ (1 - {\gamma}^5) \hat{q} (\hat{p}_3 + m
\hat{n}_3) (m - \hat{p}_4 \hat{n}_4) \right] \Tr \left[ (1 -
{\gamma}^5) \hat{q} (m + \hat{p}_2 \hat{n}_2) (\hat{p}_1 + m
\hat{n}_1) \right]
             \\[0.5cm]    \displaystyle
\cdot { \Tr \left[ ( 1 - {\gamma}^5 ) \hat{q} \hat{p}_4 \hat{n}_4
\hat{q} \hat{p}_2 \hat{n}_2 \right] \over 8 \left[ (q p_2) - m (q
n_2) \right] \cdot \left[ (q p_4) - m (q n_4)  \right] } \; .
\end{array}
\label{e2.20}
\end{equation}

The final equation is
\begin{equation}
\begin{array}{r} \displaystyle
\bar{u}_3 \gamma_{\rho} u_1 \cdot \bar{u}_4 \gamma^{\rho} u_2 = {1
\over  32 \sqrt{ (q p_1) + m (q n_1) }  \sqrt{ (q p_2) + m (q n_2)
} \sqrt{ (q p_3) + m (q n_3) } \sqrt{ (q p_4) + m (q n_4) } }
               \\[0.8cm]    \displaystyle
\cdot \Big\{ \Tr \left[ (1 - {\gamma}^5) \hat{q} (\hat{p}_3 + m
\hat{n}_3) (m + \hat{p}_2 \hat{n}_2) \right] \Tr \left[ (1 -
{\gamma}^5) \hat{q} (m - \hat{p}_4 \hat{n}_4) (\hat{p}_1 + m
\hat{n}_1) \right]
               \\[0.5cm]    \displaystyle
+ \Tr \left[ (1 - {\gamma}^5) \hat{q} (\hat{p}_4 + m \hat{n}_4) (m
+ \hat{p}_1 \hat{n}_1) \right] \Tr \left[ (1 - {\gamma}^5) \hat{q}
(m - \hat{p}_3 \hat{n}_3) (\hat{p}_2 + m \hat{n}_2) \right]
              \\[0.5cm]    \displaystyle
- \{ \Tr \left[ (1 - {\gamma}^5) \hat{q} (\hat{p}_3 + m \hat{n}_3)
(m - \hat{p}_4 \hat{n}_4) \right] \Tr \left[ (1 - {\gamma}^5)
\hat{q} (m + \hat{p}_2 \hat{n}_2) (\hat{p}_1 + m \hat{n}_1)
\right]
             \\[0.5cm]    \displaystyle
+ \Tr \left[ (1 - {\gamma}^5) \hat{q} (\hat{p}_2 - m \hat{n}_2) (m
+ \hat{p}_1 \hat{n}_1) \right] \Tr \left[ (1 - {\gamma}^5) \hat{q}
(m - \hat{p}_3 \hat{n}_3) (\hat{p}_4 - m \hat{n}_4) \right] \}
             \\[0.5cm]    \displaystyle
\cdot { \Tr \left[ ( 1 - {\gamma}^5 ) \hat{q} \hat{p}_4 \hat{n}_4
\hat{q} \hat{p}_2 \hat{n}_2 \right] \over 8 \left[ (q p_2) - m (q
n_2) \right] \cdot \left[ (q p_4) - m (q n_4)  \right] }
 \Big\} \; .
\end{array}
\label{e2.21}
\end{equation}

The expression for
$$ \displaystyle
\bar{u}_4 \gamma_{\rho} u_1 \cdot \bar{u}_3 \gamma^{\rho} u_2
$$
is obtained from (\ref{e2.21}) by replacing the subscripts $3
\leftrightarrow 4$.

The obtained formulae are simple enough and may be used for
numerical calculation of the amplitudes.

If the denominators in (\ref{e2.21}) become zero for some values
of the vectors of problem, then one needs to change the values of
arbitrary four-vector $q$ in all formulae.


\section{Construction of polarization vectors of photons from the vectors of the problem}
\setcounter{equation}{0}

If photons participate in reaction, we must to construct their
polarization vectors from other vectors coming in amplitude
calculation. It is necessary to use Gram -- Schmidt
orthonormalization process to do this (see \cite{r3.1}).

As it shown in \cite{r3.2}, in the Minkowski space, this process
give
\begin{enumerate}
\item[1.]
Let $p$ is an arbitrary four-momentum, such that
$$ \displaystyle p^2 = m^2 \ne 0 \; , $$
$a$, $b$ and $c$ are arbitrary vectors. Then four vectors $l_0$,
$l_1$, $l_2$, $l_3$ form an orthonormal basis:
\begin{equation}
\displaystyle l_0 = { p \over m } \; ,
\label{e3.1}
\end{equation}
\begin{equation}
\begin{array}{l} \displaystyle
(l_1)_{\rho} = { G\pmatrix{p &  a \\
                             p & {}_{\rho} }
 \over m
\Biggl[ -
 G\pmatrix{p & a  \\
           p & a   }
\Biggr]^{1/2} } = { m^2 a_{\rho} - (pa) p_{\rho} \over m \sqrt{ (p
a)^2 - m^2 a^2 }  } \; ,
\end{array}
\label{e3.2}
\end{equation}
\begin{equation}
\begin{array}{l} \displaystyle
(l_2)_{\rho} = - { G\pmatrix{p & a & b \\
                           p & a & {}_{\rho}  }
 \over
\Biggl[ -
 G\pmatrix{p & a  \\
           p & a   }
 G\pmatrix{p & a & b  \\
           p & a & b   }
\Biggr]^{1/2} } =
        \\[0.5cm] \displaystyle
= { \left[ a^2 (p b) - (p a) (a b) \right] p_{\rho}
  + \left[ m^2 (a b) - (p a) (p b) \right] a_{\rho}
  + \left[ (p a)^2 - m^2 a^2 \right] b_{\rho}
\over
 \sqrt{ (p a)^2 - m^2 a^2 }
 \sqrt{ 2(p a) (p b) (a b) + m^2 a^2 b^2 -
 m^2 (a b)^2 - a^2 (p b)^2 - b^2 (p a)^2 } }
\; ,
\end{array}
\label{e3.3}
\end{equation}
$$
\begin{array}{l} \displaystyle
(l_3)_{\rho}  =  { G\pmatrix{p & a & b & c\\ p & a & b & {}_{\rho}
} \over \left[- G\pmatrix{p & a & b  \\ p & a & b  } G\pmatrix{p &
a & b & c \\ p & a & b & c } \right]^{1/2} } \; ,
\end{array}
$$
where $G$ are Gram determinants. Finally we take into account that
in the Minkowski space
$$
\begin{array}{l} \displaystyle
G\pmatrix{p & a & b & c \\ p & a & b & c } = - {\varepsilon}_{
{\mu} {\nu} {\sigma} {\tau} } p^{\mu} a^{\nu} b^{\sigma} c^{\tau}
 {\varepsilon}_{
{\alpha} {\beta} {\lambda} {\kappa}} p^{\alpha} a^{\beta}
b^{\lambda} c^{\kappa} \; ,
\end{array}
$$
$$
\begin {array}{l} \displaystyle
G\pmatrix{p & a & b & c \\ p & a & b & {}_{\rho} } =
{\varepsilon}_{ {\mu} {\nu} {\sigma} {\tau} } p^{\mu} a^{\nu}
b^{\sigma} c^{\tau} {\varepsilon}_{ {\rho} {\alpha} {\beta}
{\lambda} } p^{\alpha} a^{\beta} b^{\lambda} \; .
\end {array}
$$
and last vector becomes:
\begin{equation}
\begin{array}{l} \displaystyle
(l_3)_{\rho} = { {\varepsilon}_{ {\rho} {\alpha} {\beta} {\lambda}
}  p^{\alpha} a^{\beta} b^{\lambda}
       \over
\Biggl[
 G\pmatrix{p & a & b  \\
           p & a & b   }
\Biggr]^{1/2} } = { {\varepsilon}_{ {\rho} {\alpha} {\beta}
{\lambda} } p^{\alpha} a^{\beta} b^{\lambda} \over
 \sqrt{ 2(p a) (p b) (a b) + m^2 a^2 b^2 -
 m^2 (a b)^2 - a^2 (p b)^2 - b^2 (p a)^2 } }
\; .
\end{array}
\label{e3.4}
\end{equation}
Note that in the final formulae  the vector $c$ is absent.

So, using the three vectors (one of which is a four-momentum of a
massive particle) and a total antisymmetric Levi-Civita tensor, an
orthonormal basis can be always constructed in the Minkowski
space.

\item[2.]
Let us consider now the construction of the basis for reaction
with massless particles.

Let $k_1$ and $k_2$ are four-momentums of the problem, such that
$$ \displaystyle k_1^2 = k_2^2 = 0 \, . $$
We consider a particular form of the basis
(\ref{e3.1})~--~(\ref{e3.4}) at
$$ p =  k_1 +  k_2 \, , \; m = \sqrt { 2 ( k_1 k_2 ) } \, , \; a =
k_2 \, : $$
\begin{equation}
\displaystyle
l_0 = { k_1 + k_2 \over  \sqrt{ 2 (k_1 k_2) } } \; ,
\label{e3.5}
\end{equation}
\begin{equation}
\displaystyle
l_1 = { - k_1 + k_2 \over \sqrt{ 2 (k_1 k_2) } } \; ,
\label{e3.6}
\end{equation}
\begin{equation}
\displaystyle
(l_2)_{\rho} = { - (k_2 b) (k_1)_{\rho} - (k_1 b) (k_2)_{\rho}  +
(k_1 k_2) b_{\rho} \over \sqrt{ 2(k_1 k_2) (k_1 b) (k_2 b) - b^2
(k_1 k_2)^2 } } \; ,
\label{e3.7}
\end{equation}
\begin{equation}
\displaystyle
(l_3)_{\rho} = { {\varepsilon}_{ {\rho} {\alpha} {\beta} {\lambda}
}
     k_1^{\alpha} k_2^{\beta} b^{\lambda}
 \over
 \sqrt{ 2(k_1 k_2) (k_1 b) (k_2 b)  - b^2 (k_1 k_2)^2 } }
\; .
\label{e3.8}
\end{equation}

\end{enumerate}

If $ \displaystyle k_1 $ is four-momentum of photon, polarizations
vectors for it may be constructed from (\ref{e3.7}), (\ref{e3.8}),
where it is appropriate to choose
$ \displaystyle b = q $,
[$q$ is the same massless vector which contained in expressions
for amplitudes (\ref{e2.3}), (\ref{e2.4})].
$$ \displaystyle (l_2)_{\rho} = { - (k_2 q) {k_1}_{\rho} + (k_1
k_2) q_{\rho}  - (k_1 q) {k_2}_{\rho} \over
 \sqrt{ 2(k_1 k_2) (k_1 q) (k_2 q) } }
\; , $$
$$ \displaystyle (l_3)_{\rho} = - { {\varepsilon}_{ {\rho}
{\alpha} {\beta} {\lambda} }
     k_1^{\alpha} q^{\beta} k_2^{\lambda}
 \over  \sqrt{ 2(k_1 k_2) (k_1 q) (k_2 q) } }
 \; , $$
\begin{equation}
\displaystyle
e^{\pm}_{\rho}(k_1) = - { (l_2)_{\rho} \pm  i (l_3)_{\rho} \over
\sqrt{2} } = { \Tr \left[ ( 1 \pm {\gamma}^5) {\gamma}_{\rho}
{\hat k_1} {\hat q} {\hat k_2} \right]
 \over 8  \sqrt{ (k_1 k_2) (k_1 q) (k_2 q) } }
 \; .
\label{e3.9}
\end{equation}
It is easy to check that this vector satisfies standard conditions
for polarization vector:
\begin{equation}
\displaystyle
(k_1 e^{\pm} ) = (e^{\pm})^2 = 0 , \, (e^{\pm} e^{\mp}) = -1 \; .
\label{e3.10}
\end{equation}

Then, using formulae of Fierz transformation
$$ \displaystyle
 \left[ ( 1 \mp \gamma^5 ) \gamma^{\rho} \right]_{kl}
 \left[ ( 1 \pm \gamma^5 ) \gamma_{\rho} \right]_{ij}
 = 2 [ 1 \mp \gamma^5 ]_{kj} [ 1 \pm \gamma^5 ]_{il}  \, , $$
$$ \displaystyle
 \left[ (1 \pm \gamma^5 ) \gamma^{\rho} \right]_{kl}
 \left[ (1 \pm \gamma^5 ) \gamma_{\rho} \right]_{ij}
= - \left[ ( 1 \pm \gamma^5 ) \gamma^{\rho} \right]_{kj}
    \left[ ( 1 \pm \gamma^5 ) \gamma_{\rho} \right]_{il}
\, , $$
($i$, $j$, $k$, $l$ are indices that label the components  of $4
\times 4$-matrices), we obtain:
$$ \displaystyle
 ( 1 \mp \gamma^5 ) \gamma^{\rho}
 \Tr \left[ ( 1 \pm {\gamma}^5) {\gamma}_{\rho} {\hat k}_1 {\hat
q} {\hat k}_2 \right] = 2 ( 1 \mp \gamma^5 ) {\hat k}_1 {\hat q}
{\hat k}_2 ( 1 \pm {\gamma}^5) = 4 ( 1 \mp \gamma^5 ) {\hat k}_1
{\hat q} {\hat k_2} \, ,
$$
$$
\begin {array}{l} \displaystyle
( 1 \pm \gamma^5 ) \gamma^{\rho}
 \Tr \left[ ( 1 \pm {\gamma}^5) {\gamma}_{\rho} {\hat k}_1 {\hat
q} {\hat k}_2 \right] = -  ( 1 \pm \gamma^5 )  \gamma^{\rho} {\hat
k}_1 {\hat q} {\hat k}_2 ( 1 \pm \gamma^5 ) {\gamma}_{\rho} =
        \\[0.5cm] \displaystyle
= - 2 ( 1 \pm \gamma^5 )  \gamma^{\rho} {\hat k}_1 {\hat q} {\hat
k}_2 {\gamma}_{\rho} = 4 ( 1 \pm \gamma^5 ) {\hat k}_2 {\hat q}
{\hat k}_1
 \, .
\end {array}
$$
Therefore
$$ \displaystyle
 \gamma^{\rho}
 \Tr \left[ ( 1 \pm {\gamma}^5) {\gamma}_{\rho} {\hat k}_1 {\hat
q} {\hat k}_2 \right] = 2 ( 1 \mp \gamma^5 ) {\hat k}_1 {\hat q}
{\hat k}_2 + 2 ( 1 \pm \gamma^5 ) {\hat k}_2 {\hat q} {\hat k}_1
\,  $$
and we have
\begin{equation}
\displaystyle
{\hat e}^{\pm} (k_1) = { ( 1 \pm {\gamma}^5 ) {\hat k}_2 {\hat q}
{\hat k}_1 + ( 1 \mp {\gamma}^5 ) {\hat k}_1 {\hat q} {\hat k}_2
 \over
4 \sqrt{ (k_1 k_2) (k_1 q) (k_2 q) } } \; .
\label{e3.11}
\end{equation}

Note that $\displaystyle k_2$ is an arbitrary massless vector of
problem. It follows from this that for photon with four-momentum
$\displaystyle k_2$ polarization vectors may be chosen in form
(\ref{e3.9}) too (see also \cite{r3.3}):
\begin{equation}
\displaystyle
e^{\pm}_{\rho}(k_1) = { \Tr \left[ ( 1 \pm {\gamma}^5)
{\gamma}_{\rho} {\hat k_1} {\hat q} {\hat k_2} \right]
 \over 8  \sqrt{ (k_1 k_2) (k_1 q) (k_2 q) } }
 = { \Tr \left[ ( 1 \mp {\gamma}^5) {\gamma}_{\rho}
{\hat k_2} {\hat q} {\hat k_1} \right]
 \over 8  \sqrt{ (k_1 k_2) (k_1 q) (k_2 q) } } = e^{\mp}_{\rho}(k_2)
 \; .
\label{e3.12}
\end{equation}
This circumstance simplifies calculations for two-photon process
essentially. For example, take place expression
\begin{equation}
\displaystyle
e^{\pm}_{\mu} (k_1) e^{\pm}_{\nu} (k_2) = { \Tr \left[ ( 1 \pm
{\gamma}^5) {\gamma}_{\mu} {\hat k}_1 {\gamma}_{\nu} {\hat k}_2
\right]
 \over
8 (k_1 k_2) } \; ,
\label{e3.13}
\end{equation}
et al.

In conclusion of this section we consider one more problem. In
reality, polarization vectors for the photon with ñ four-momentum
$k_1$, satisfied (\ref{e3.10}), may be constructed with help of
any vectors $a$ and $b$:
\begin{equation}
\begin{array}{l} \displaystyle
e^{\pm}_{\rho}(k_1; a,b) = { \Tr \left[ ( 1 \pm {\gamma}^5)
{\gamma}_{\rho} {\hat k_1} {\hat a} {\hat b} \right] \over 4
\sqrt{ 2 G\pmatrix{k_1 & a & b \\ k_1 & a & b } } } \; .
\end{array}
\label{e3.14}
\end{equation}
Using identity (\ref{e2.6}) we have
$$\begin {array}{l} \displaystyle
\Tr [ ( 1 \pm \gamma^5 ) \gamma_{\rho} \hat{k}_1 \hat{a} \hat{b} ]
\Tr [ ( 1 \mp \gamma^5 ) \hat{k}_1 \hat{d} \hat{c} \hat{k}_1
\hat{c} \hat{d}] = \Tr [ ( 1 \pm \gamma^5 ) \gamma_{\rho}
\hat{k}_1 \hat{a} \hat{b} ] 8 G\pmatrix{k_1 & c & d  \\ k_1 & c &
d}
=
        \\[0.5cm] \displaystyle
= \Tr [ ( 1 \mp \gamma^5 ) \hat{k}_1 \hat{a} \hat{b} \gamma_{\rho}
( 1 \mp \gamma^5 ) \hat{k}_1 \hat{d} \hat{c} \hat{k}_1 \hat{c}
\hat{d}] = \Tr [ \hat{k}_1 \hat{a} \hat{b}( 1 \pm \gamma^5 )
\gamma_{\rho} \hat{k}_1 \hat{d} \hat{c} ( 1 \pm \gamma^5 )
\hat{k}_1 \hat{c} \hat{d}] =
          \\[0.5cm] \displaystyle
= 4 {k_1}_{\rho} \Tr [ ( 1 \mp \gamma^5 )\hat{k}_1 \hat{a} \hat{b}
\hat{d} \hat{c} \hat{k}_1 \hat{c} \hat{d}] - \Tr [ \hat{k}_1
\hat{a} \hat{b}( 1 \pm \gamma^5 ) \hat{k}_1 \gamma_{\rho} \hat{d}
\hat{c} ( 1 \pm \gamma^5 ) \hat{k}_1 \hat{c} \hat{d}] =
             \\[0.5cm] \displaystyle
= 4 {k_1}_{\rho} \Tr [ ( 1 \mp \gamma^5 )\hat{k}_1 \hat{a} \hat{b}
\hat{d} \hat{c} \hat{k}_1 \hat{c} \hat{d}] - \Tr [( 1 \pm \gamma^5
) \gamma_{\rho} \hat{k}_1 \hat{c} \hat{d} ] \ \Tr [( 1 \pm
\gamma^5 ) \hat{k}_1 \hat{c} \hat{d} \hat{k}_1 \hat{a} \hat{b}] \;
.
\end {array}  $$
Thus
$$\begin {array}{l} \displaystyle
e^{\pm}_{\rho}(k_1; a,b) = - e^{\pm}_{\rho}(k_1; c,d) { \Tr [ ( 1
\pm \gamma^5 ) \hat{k}_1 \hat{c} \hat{d} \hat{k}_1 \hat{a}
\hat{b}] \over 8 \sqrt{ G\pmatrix{k_1 & a & b  \\ k_1 & a & b}
G\pmatrix{k_1 & c & d
\\ k_1 & c & d} } } +
   \\[0.7cm] \displaystyle
+ {k_1}_{\rho} { \Tr [ ( 1 \mp \gamma^5 ) \hat{k}_1 \hat{a}
\hat{b} \hat{d} \hat{c} \hat{k}_1 \hat{c} \hat{d}]  \over \sqrt{ 2
G\pmatrix{k_1 & a & b  \\ k_1 & a & b} } \cdot 8G\pmatrix{k_1 & c
& d
\\ k_1 & c & d} }  \; .
\end {array} $$
Since last term may be neglected because of gauge invariance, we
obtain that replacement of the vectors $a$, $b$ in polarization
vector by another vectors $c$, $d$ leads to the phase factor
$$\begin {array}{l} \displaystyle
- { \Tr [ ( 1 \pm \gamma^5 ) \hat{k}_1 \hat{c} \hat{d} \hat{k}_1
\hat{a} \hat{b}]  \over  8 \sqrt{ G\pmatrix{k_1 & a & b  \\ k_1 &
a & b} G\pmatrix{k_1 & c & d  \\ k_1 & c & d} } } = - \left(
e^{\pm}(k_1;a,b) e^{\mp}(k_1;c,d) \right) \; .
\end {array} $$
%


\section{Some remarks about other methods of amplitude calculation}
\setcounter{equation}{0}

Formula (\ref{e1.10}) is a particular case of general equation:
\begin{equation}
\displaystyle
\bar{u}_f Q u_i \simeq ( \bar{u}_f Q u_i ) \cdot
       { \bar{u}_i Z u_f \over |\bar{u}_i Z u_f| }
= { \Tr ( Q u_i \bar{u}_i Z u_f \bar{u}_f ) \over
         | \bar{u}_i Z u_f | }
     = { \Tr ( Q u_i \bar{u}_i Z u_f \bar{u}_f ) \over
\sqrt{ \Tr ( \bar{Z} u_i \bar{u}_i Z u_f \bar{u}_f ) } }
\label{e4.1}
\end{equation}
where $Z$ is matrix operator,
$ \displaystyle \bar{Z} = \gamma^0 Z^{+} \gamma^0 $.
Formula (\ref{e4.1}) describes all methods of amplitude
calculation where $Z$ may be
$ \displaystyle 1  \; ,  \gamma^5  \; , \gamma^0 \;, 1+\gamma^0 \;
$
et al. Detailed classification of different methods is given in
\cite{r1.2}. Many methods are analyzed in the book \cite{r4.1}
also.

The most of the methods give more simple expressions for
amplitudes in comparison with (\ref{e1.10}). For example, if
$ \displaystyle Z = 1$
(see\cite{r4.2}), we have:
\begin{equation}
\displaystyle
\bar{u}_f Q u_i \simeq { \Tr \left[ (\hat{p}_f + m_f)( 1 +
\gamma^5 \hat{n}_f) Q (\hat{p}_i + m_i)( 1 + \gamma^5 \hat{n}_i)
\right] \over 4 \sqrt{ \left[ m_i m_f + ( p_i p_f ) \right] \left[
1 - ( n_i n_f ) \right] + ( p_i n_f ) ( p_f n_i )  } } \; .
\label{e4.2}
\end{equation}

However, the scheme considered here has two essential
disadvantages in general case:
\begin{description}
\item[1.]
There is denominator in (\ref{e4.1}), hence the ambiguity of the
type
$\displaystyle {0 \over 0} $
can appear during the calculations. For example, in (\ref{e4.2})
the denominator is equal to zero for
$$
\displaystyle
n_f = - n_i + { ( p_f n_i ) \over
                    m_i m_f + ( p_i p_f ) }
  ( p_i + { m_i \over m_f } p_f ) \; .
$$
\item[2.]
Each of the fermion lines obtain phase factor [see (\ref{e4.1})]:
$$\displaystyle
{ \bar{u}_i Z u_f \over  | \bar{u}_i Z u_f | } \;\; .
$$
If we have the interference diagrams, this circumstance leads to
the fact that expressions for the amplitudes corresponding to
different channels obtain different phase factors in general case.
In this situation the additional calculations are necessary:
calculation of the relative phase for interference diagrams, the
Fierz transformations et al. (see \cite{r1.2}).
\end{description}
At the same time, the presented method, as mentioned above, has no
these disadvantages.



\end {document}